\documentclass[a4paper]{PoS}

\usepackage{amsmath}
\usepackage{amssymb}
\usepackage{graphicx}
\usepackage{cite}
\usepackage{multirow}

\title{
	Conformal symmetry vs. chiral symmetry breaking in the SU(3) sextet model
}

\ShortTitle{
	Conformal symmetry vs. chiral symmetry breaking in the SU(3) sextet model
}

\author{
	Vincent Drach \\
	CP$^3$-Origins and the Danish IAS, University of Southern Denmark \\
	E-mail: \email{drach@cp3-origins.net}
}

\author{
	\speaker{Martin Hansen} \\
	CP$^3$-Origins and the Danish IAS, University of Southern Denmark \\
	E-mail: \email{hansen@cp3-origins.net}
}

\author{
	Ari Hietanen \\
	CP$^3$-Origins and the Danish IAS, University of Southern Denmark \\
	E-mail: \email{hietanen@cp3-origins.net}
}

\author{
	Claudio Pica \\
	CP$^3$-Origins and the Danish IAS, University of Southern Denmark \\
	E-mail: \email{pica@cp3-origins.net}
}

\author{
	Francesco Sannino \\
	CP$^3$-Origins and the Danish IAS, University of Southern Denmark \\
	E-mail: \email{sannino@cp3-origins.net}
}

\abstract{
We present new results for the SU(3) ``sextet model'' with two flavors transforming according to the two-index symmetric representation of the gauge group. The simulations are performed using unimproved Wilson fermions. We measure the meson and baryon spectrum of the theory for multiple bare quark masses at two different lattice spacings. To address the pressing issue of whether the model is inside or below the conformal window, we compare the spectrum to the expectations for a theory with spontaneous chiral symmetry breaking and to those of an IR conformal theory. Regardless of the answer (conformal or chirally broken), the theory is a cornerstone in our understanding of near-conformal and composite dynamics, ranging from Technicolor models to unparticle physics. It is also interesting for  the composite dynamics of vector-like singlets with respect to the Standard Model interactions. 
\\[2mm]
\it Preprint: CP$^3$-Origins-2015-034 DNRF90, DIAS-2015-34
}

\FullConference{
	The 33rd International Symposium on Lattice Field Theory \\
	14 - 18 July 2015 \\
	Kobe International Conference Center, Kobe, Japan
}

\begin{document}

\section{Introduction}
\vspace{-3mm}

The discovery of a new scalar state, in agreement with the theoretical predictions for the Higgs boson, is a tremendous success for the Large Hadron Collider and for the Standard Model. Nevertheless, from a theoretical standpoint the Higgs sector is quite unappealing. Partly because it only models spontaneous symmetry breaking, rather than offering a dynamical explanation, but mostly because it does not protect the electroweak scale against quantum corrections. The latter is known as the naturalness problem.

The discovery of the Higgs boson excluded a large number of BSM models, such as the ``Higgsless models'', where by construction the effective Lagrangian did not include the Higgs boson. Traditional Technicolor theories, based on QCD-like physics, are also excluded because the scaled-up version of the scalar state cannot be reconciled with experiments. 

However, wide classes of composite theories are still very much in the game. The first very promising one is constituted by minimal models of fundamental near-conformal dynamics (FCD) known as Minimally Walking Technicolor (MWT) \cite{Sannino:2004qp,Dietrich:2006cm}, where the fermion condensate is responsible for breaking the electroweak symmetry and where the Higgs boson is the lightest scalar excitation of this condensate. The second class is the Composite Goldstone Higgs models, where the Higgs boson mainly is identified with one of the pseudo-Goldstone bosons of the theory \cite{Kaplan:1983sm}. Furthermore, any four-dimensional fundamental realisation of the Composite Goldstone Higgs admits a Technicolor-like limit and fine-tuning is needed to align the vacuum in the Goldstone Higgs direction \cite{Cacciapaglia:2014uja} . 

Lattice simulations provide invaluable information about the FCD, before coupling it to the EW sector \cite{Foadi:2012bb}. The MWT model studied here is an SU(3) gauge theory with a doublet of fermions in the two-index symmetric (sextet) representation. From perturbation theory this model is expected to be close to the conformal window, but still in the chirally broken phase \cite{Sannino:2004qp}. There are two prominent features of near-conformal models which make them a possible candidate for a strongly coupled Higgs sector. Because of the proximity to the conformal window a light $0^{++}$ scalar state can emerge and play the role of the Higgs boson. Additionally, because we are close to an IR fixed-point the $\beta$-function is believed to be walking. This model is furthermore expected to have a small $S$-parameter to not violate constraints on electroweak precision tests \cite{Sannino:2004qp}.

Non-perturbative calculations are required to confirm whether the model is inside or below the conformal window. If the model is chirally broken (below the conformal window), they are also needed to determine the mass of the lightest scalar state. Predictions for spin-1 resonances will furthermore be crucial for searches at the LHC. If the theory is conformal, the existence of four-fermion interactions can drive the theory away from conformality \cite{Fukano:2010yv}. The sextet model has previously been studied \cite{Fodor:2012ty,Fodor:2015eea} with staggered fermions, showing strong evidence for chiral symmetry breaking and the existence of a light scalar state. The $\beta$-function has also been calculated both using staggered fermions \cite{Fodor:2015zna} and HYP smeared Wilson fermions \cite{Hasenfratz:2015ssa}. However, the two results are in tension, with the latter result indicating the existence of an IR fixed-point.

\section{Spectrum}
\vspace{-3mm}
\begin{figure}
\begin{center}
 \includegraphics[scale=0.5]{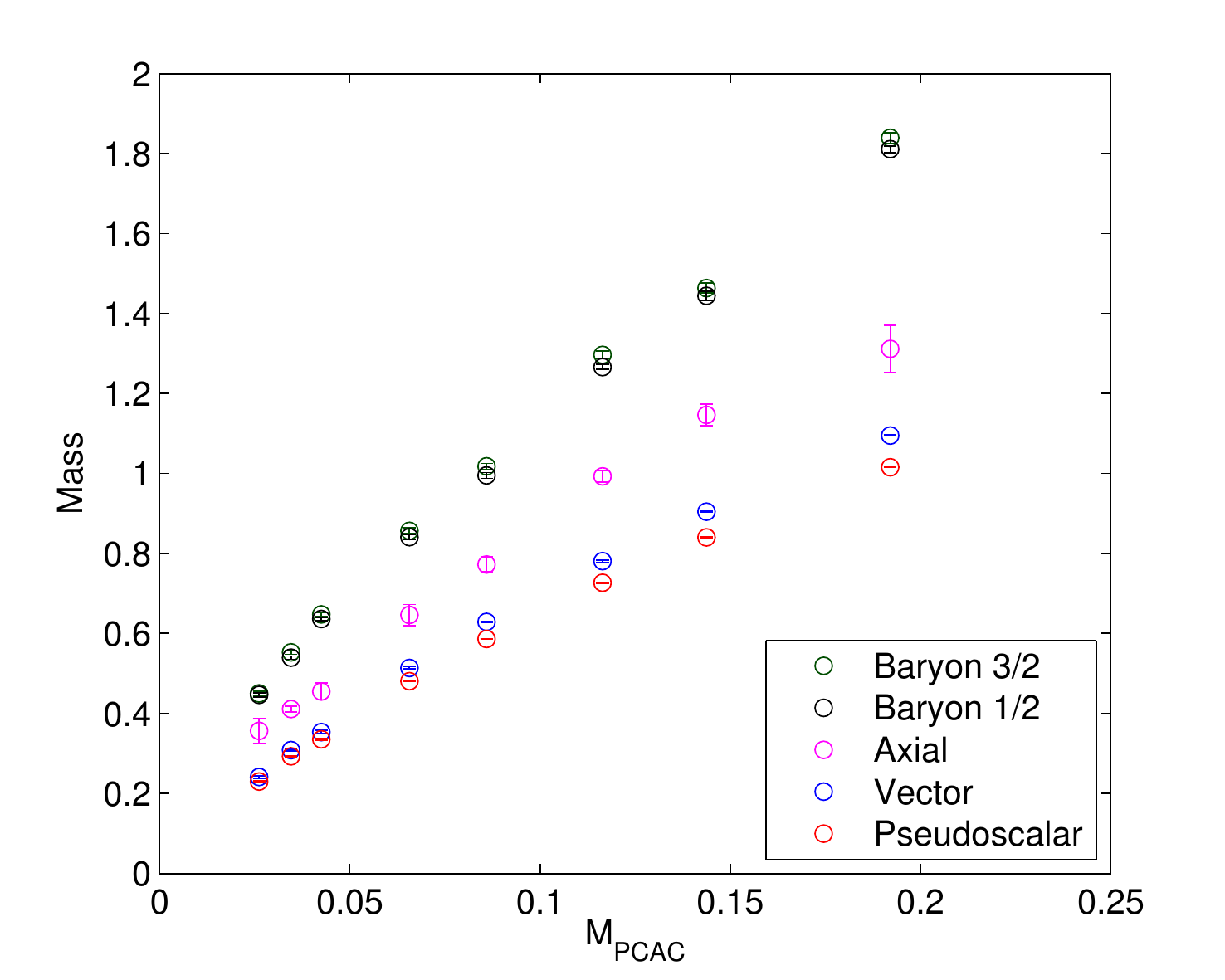}
 \includegraphics[scale=0.5]{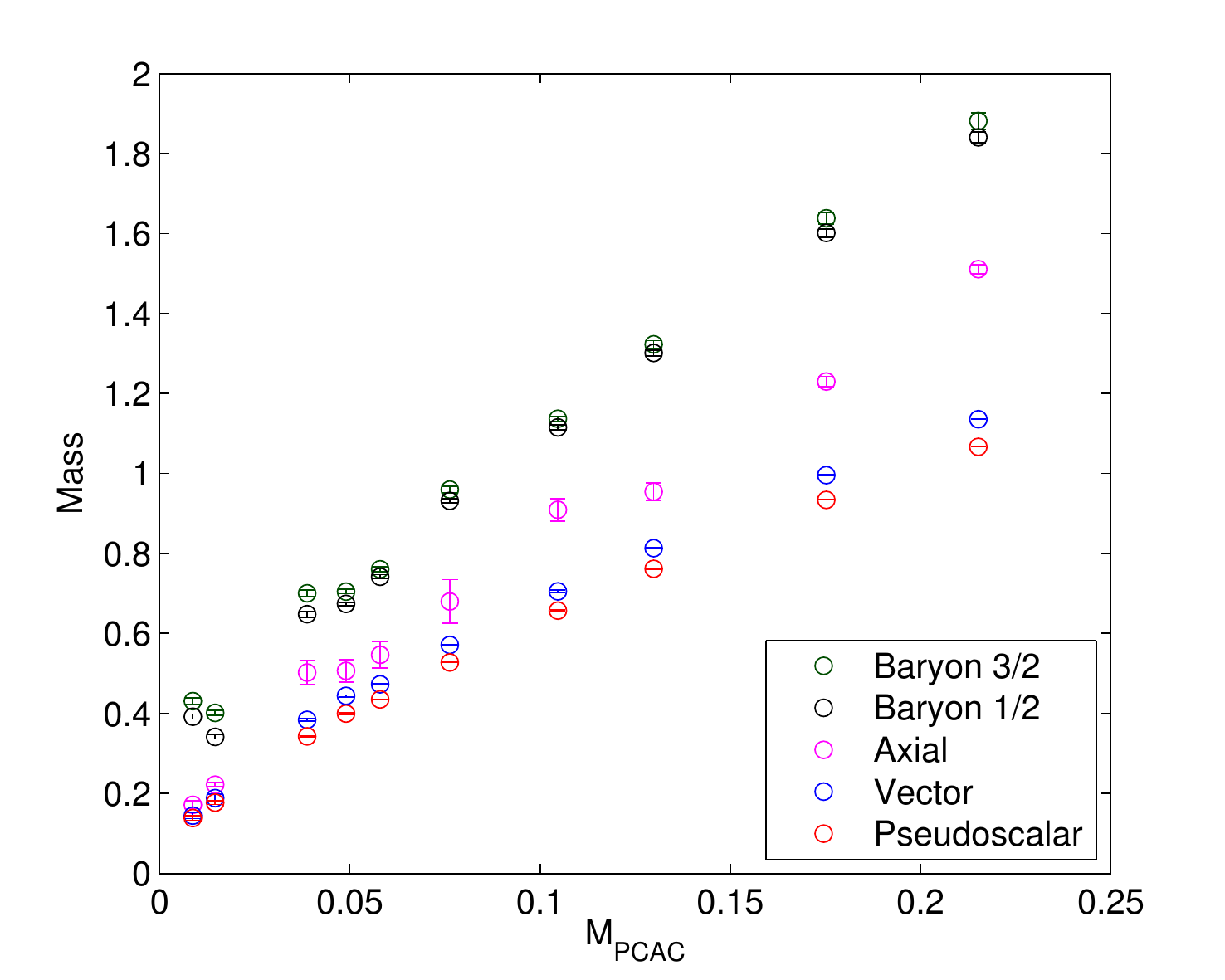}
\end{center}
\caption{Spectrum for two lattice spacings $\beta=5.4$ (left) and $\beta=5.5$ (right). The two baryon states are labeled by their spin quantum number.}
\label{fig:spectrum}
\end{figure}
In our simulations we use unimproved Wilson fermions and the plaquette gauge action. Here we present results from two different lattice spacings. At $\beta=5.4$ we have 8 different bare masses and at $\beta=5.5$ we have 10 different masses. For all masses we have around 1000 thermalised configurations. The volumes are $24^3\times40$ for the heavy masses and $32^3\times48$ for the lighter masses. We summarize the parameters for our runs in Table~\ref{tab:runs}.
\begin{table}[h]
\begin{center}
\begin{tabular}{c|c|l}
  $\beta$ & $L^3\times T$ & $-m_0$ \\
  \hline
  \multirow{2}{*}{$5.4$} & $24^3\times32$ & $1.2750,~ 1.3000,~ 1.3125,~ 1.3250,~ 1.3325,~ 1.3400$ \\
                         & $32^3\times48$ & $1.3425,~ 1.3450$ \\
  \hline
  \multirow{2}{*}{$5.5$} & $24^3\times32$ & $1.2250,~ 1.2500,~ 1.2750,~ 1.2875,~ 1.3000,~ 1.3075,~ 1.31125,~ 1.3150$ \\
                         & $32^3\times48$ & $1.3225,~ 1.3245$ \\
 \end{tabular}
 \caption{Lattice sizes and bare masses $m_0$ for our simulations.}
 \label{tab:runs}
\end{center}
\vspace{-5mm}
\end{table}
In figure \ref{fig:spectrum} we show the spectrum at $\beta=5.4$ (left) and $\beta=5.5$ (right). An interesting feature of this model is the existence of baryon spectrum analogous to that of QCD. For both lattice spacings we observe the expected hierarchy of masses, but the difference between the vector and pseudoscalar mass is quite small and the same is true for the two baryon states. At $\beta=5.5$ the runs at lighter fermion masses present larger systematic uncertainties that needs to be corrected due to e.g. the identification of plateaux in the effective masses. This is especially evident in the extraction of the baryon masses, but it is present to a lesser extent also for the axial vector channel. Due to this systematic effect, the baryons at lightest PCAC mass appear to be heavier than expected.

\section{Behaviour in the Chiral Limit}
\vspace{-3mm}

In the chiral limit, the behaviour of the spectrum depends on whether the theory is conformal or not. We will consider the two cases separately.

\subsection*{Chirally Broken}
\vspace{-2mm}
\begin{figure}
\begin{center}
 \vspace{-3mm}
 \includegraphics[width=0.48\textwidth]{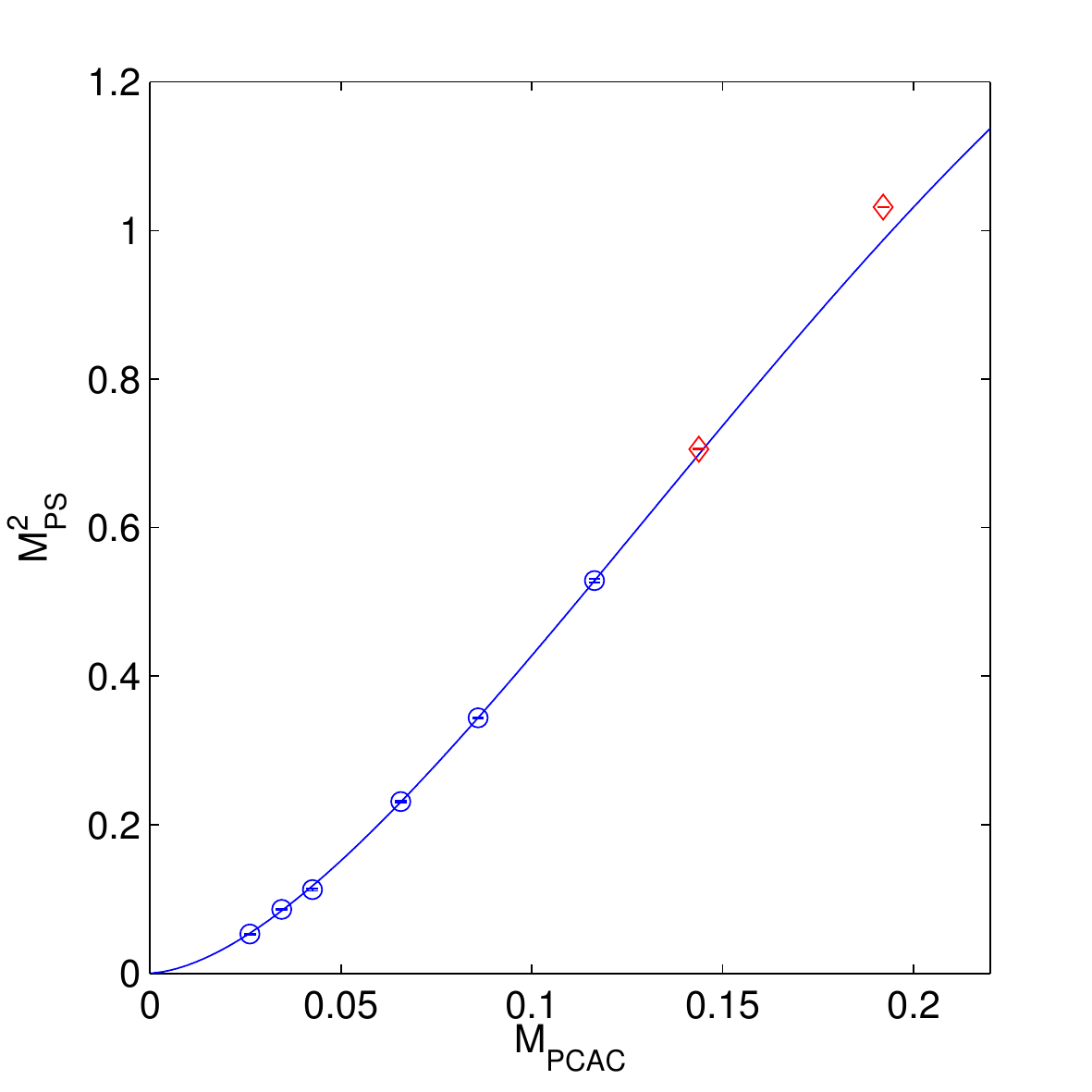}
 \includegraphics[width=0.48\textwidth]{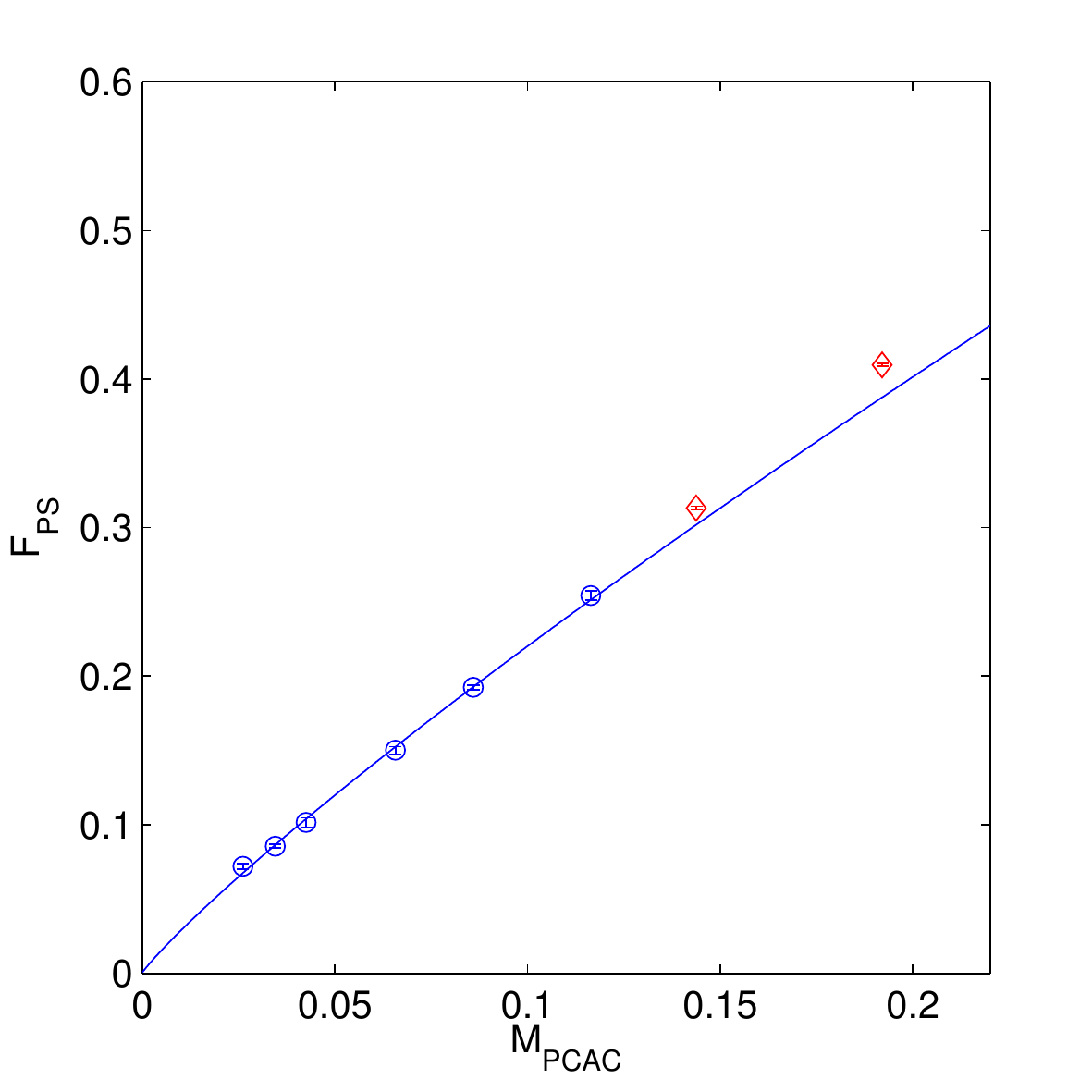} \\
 \vspace{-3mm}
 \includegraphics[width=0.48\textwidth]{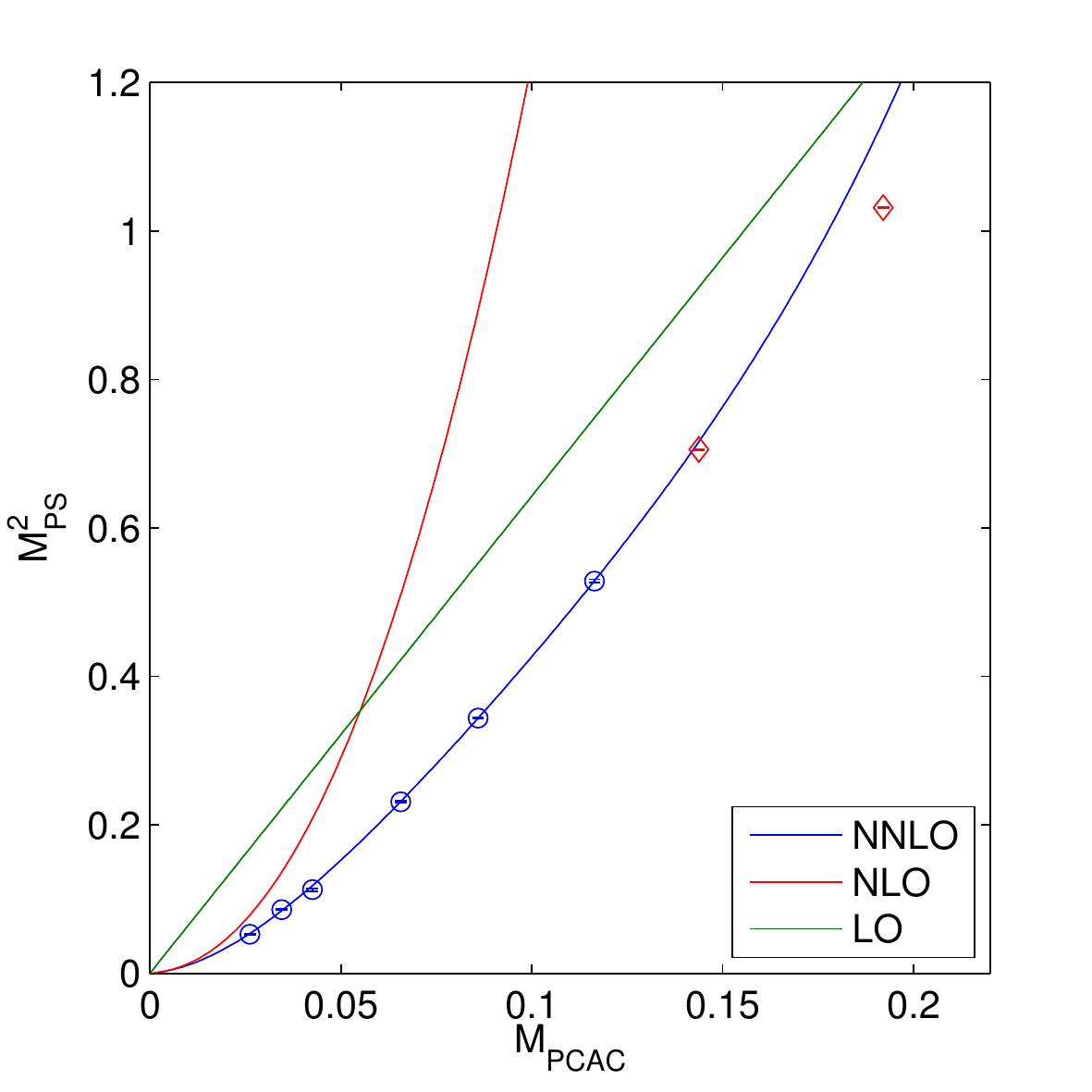}
 \includegraphics[width=0.48\textwidth]{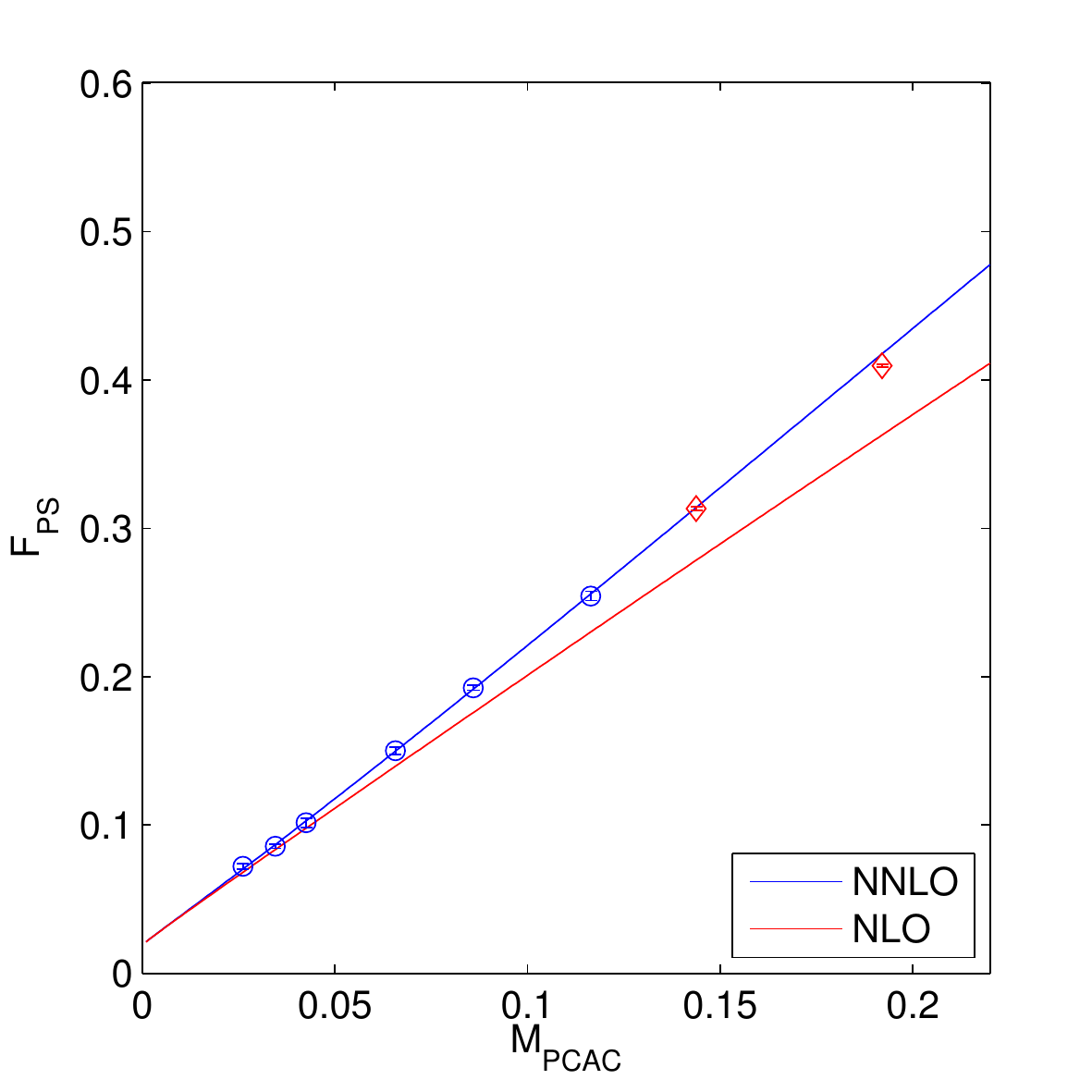} \\
 \vspace{-3mm}
\end{center}
\caption{Chiral fits for $\beta=5.4$. \textbf{Top}: Result at next-to-leading order with free coefficients for the log terms. \textbf{Bottom}: Result at next-to-next-to-leading order with continuum logs. The green and red line is the LO and NLO contributions to the NNLO fit.}
\label{fig:chpt_nlo}
\end{figure}

In a chirally broken theory the low-energy spectrum can be described by an effective Lagrangian as done in Chiral Perturbation Theory (ChPT). The behaviour of the pseudoscalar (pion) mass $M_\pi$ and decay constant $F_\pi$ can be calculated, as a function of the quark mass, to different orders in perturbation theory. At next-to-leading order (NLO) order the two quantities are given by
\begin{align}
\begin{split}
 M^2_\pi &= M^2 + \frac{M^4}{F^2}(a_ML + b_M), \\
 F_\pi &= F + \frac{M^2}{F}(a_FL + b_F).
\end{split}
\label{eq:chpt}
\end{align}
Here $M^2 = 2B_0m$ is the leading order pion mass with $m$ being the quark mass and $B_0$ being related to the chiral condensate. The full calculation at next-to-next-to-leading order (NNLO) can be found in \cite{Bijnens:2009qm}. The log-term appearing in the expressions is defined as
\begin{equation}
 L = \frac{1}{16\pi^2}\log\left(\frac{M^2}{\mu^2}\right),
\end{equation}
where $\mu$ is the reference energy scale. For the symmetry breaking pattern $SU(2) \times SU(2) \to SU(2)$ the constants in front of the log-terms are $a_M=1/2$ and $a_F=-1$ in the continuum, but they will be different at non-zero lattice spacing due to $O(a)$ corrections. The remaining constants $b_M$ and $b_F$ contain a linear combination of low-energy constants. For a chirally broken theory, equation \eqref{eq:chpt} predicts the pion mass to be zero in the chiral limit, but with a non-zero decay constant. The remaining particles in the spectrum are expected to remain massive in the chiral limit.

In figure \ref{fig:chpt_nlo} we show the chiral fits at NLO and NNLO. The blue points are the ones included in the fits. Because the continuum NLO prediction does not describe the data well, we show in the top row the result of the NLO fit with free (fitted) coefficients for the log terms. At NNLO the presence of higher order terms in the expansion makes it possible to fit the data when the log-terms are fixed to the continuum value. The NNLO result is shown in the bottom row.
\begin{table}[h]
\begin{center}
\begin{tabular}{l|cccc}
  & dofs & \# of fitted parameters & $\chi^2$/dof \\
  \hline
  NLO (free fitted logs)     & 6 & 6 & 3.5 \\
  NNLO (continuum logs) & 4 & 8 & 3.8
 \end{tabular}
 \caption{Parameters and results for the chiral fits.}
 \label{tab:chiral}
\end{center}
\vspace{-5mm}
\end{table}
In Table~\ref{tab:chiral} we show the relevant information about the chiral fits. In both cases $\chi^2$/dof is relatively large, indicating that these fits are tension with the data, as to be expected in the regime explored so far. In fact ChPT is based on the assumption that the pion is the only light particle in the spectrum. Because of the small splitting between the vector and pseudoscalar in our data, this is clearly not the case here. In addition, there might also be a light scalar state~\cite{Fodor:2012ty} making the ChPT assumption less reliable.  

\subsection*{Conformal}
\vspace{-2mm}
\begin{figure}
\begin{center}
 \vspace{-3mm}
 \includegraphics[width=0.48\textwidth]{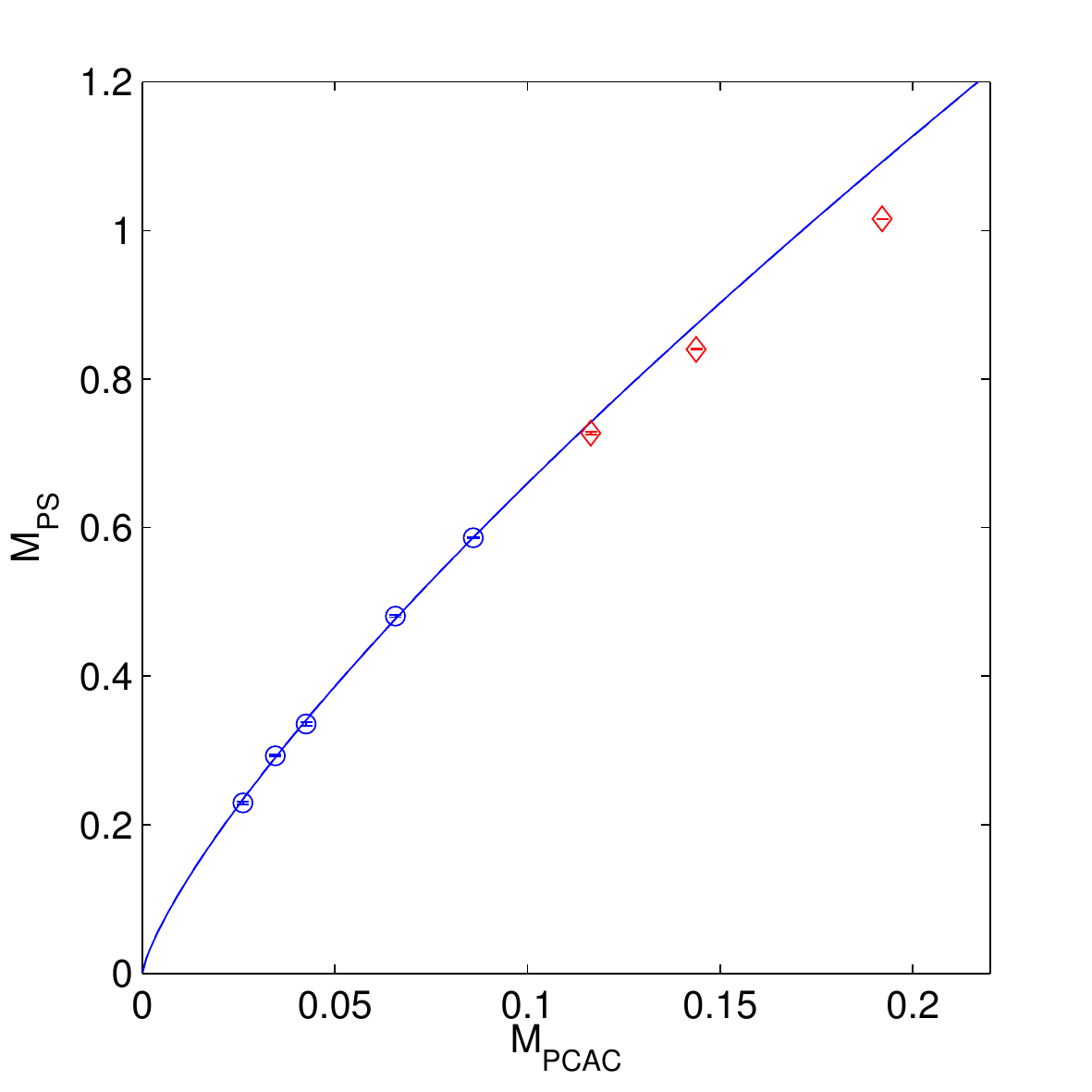}
 \includegraphics[width=0.48\textwidth]{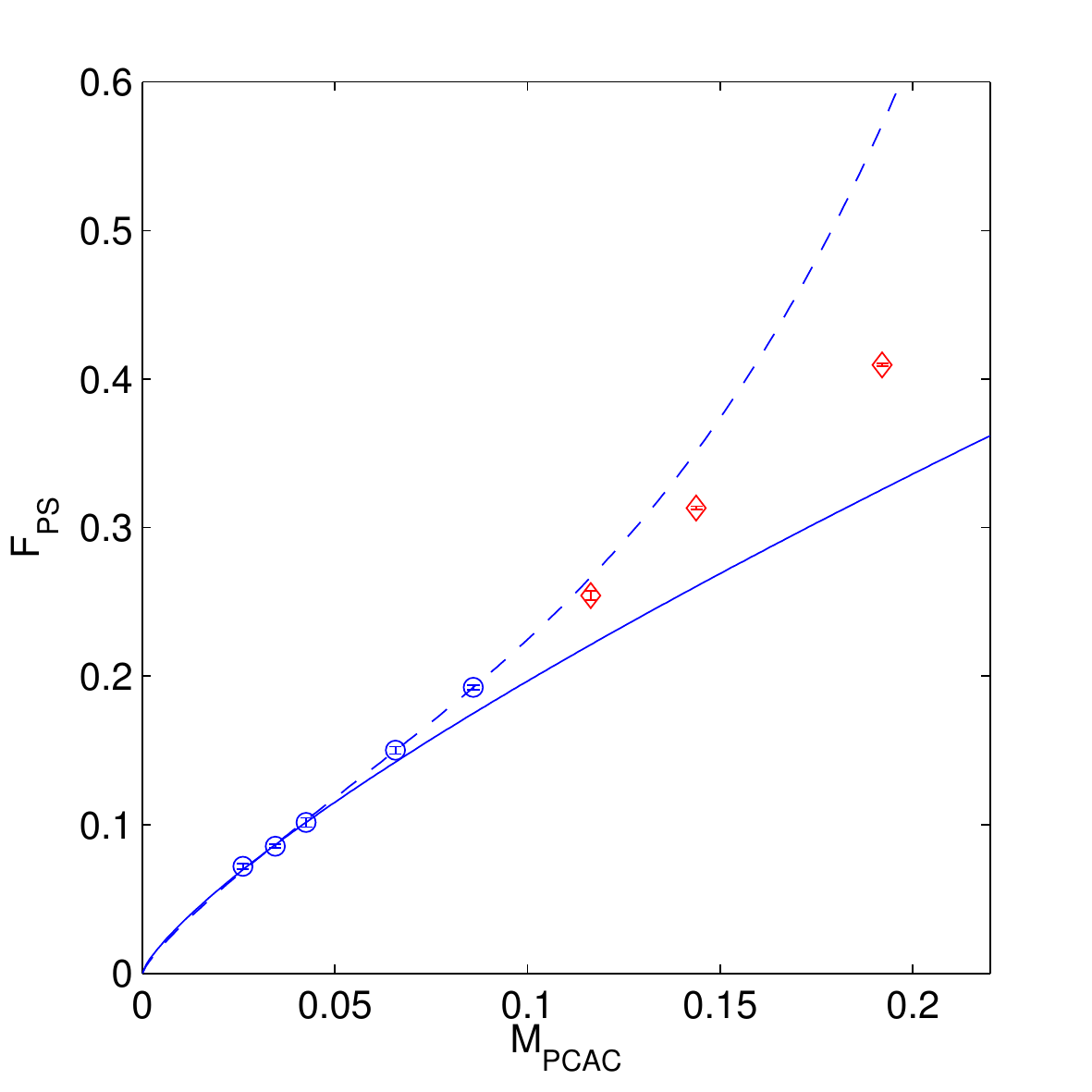} \\
 \vspace{-3mm}
 \includegraphics[width=0.48\textwidth]{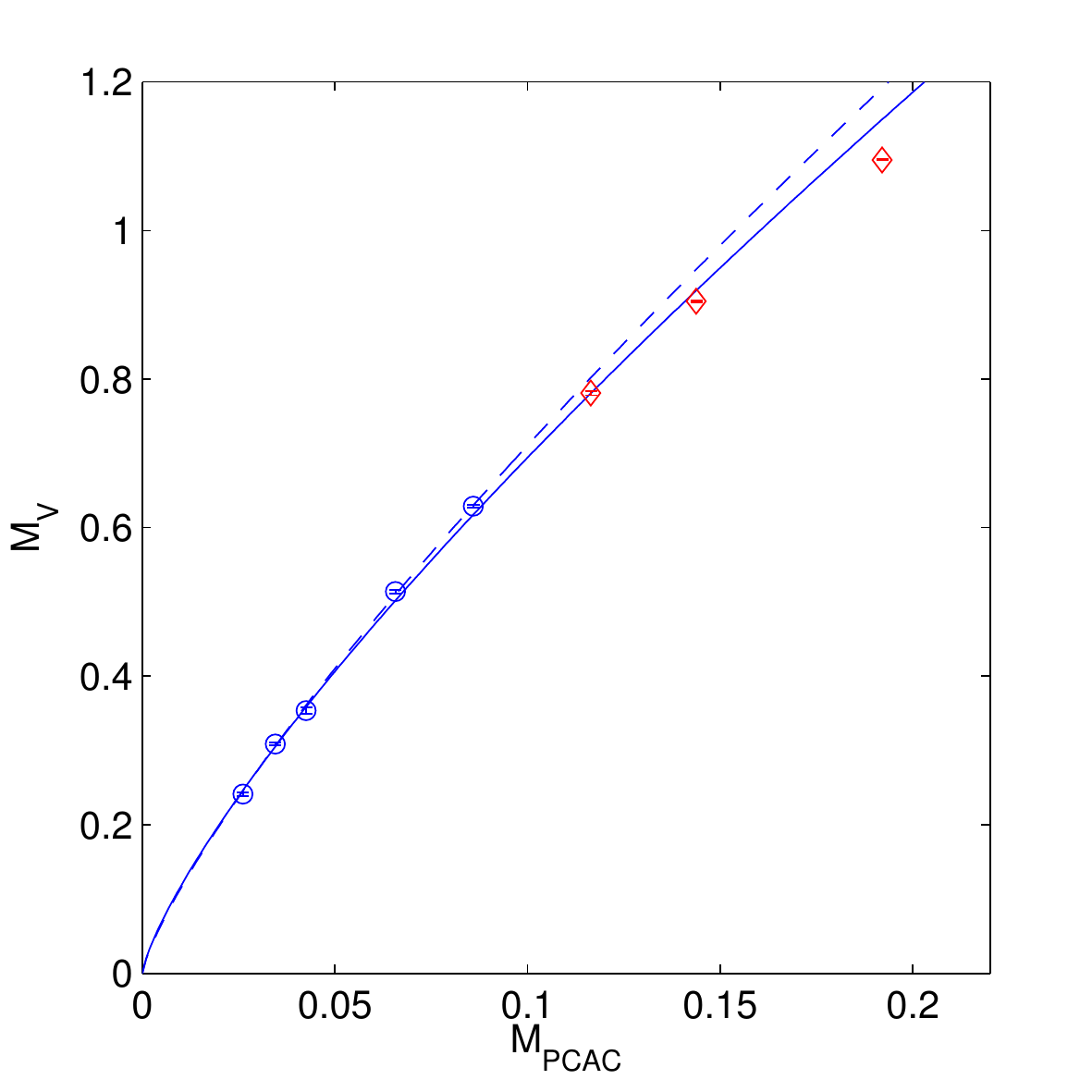}
 \includegraphics[width=0.48\textwidth]{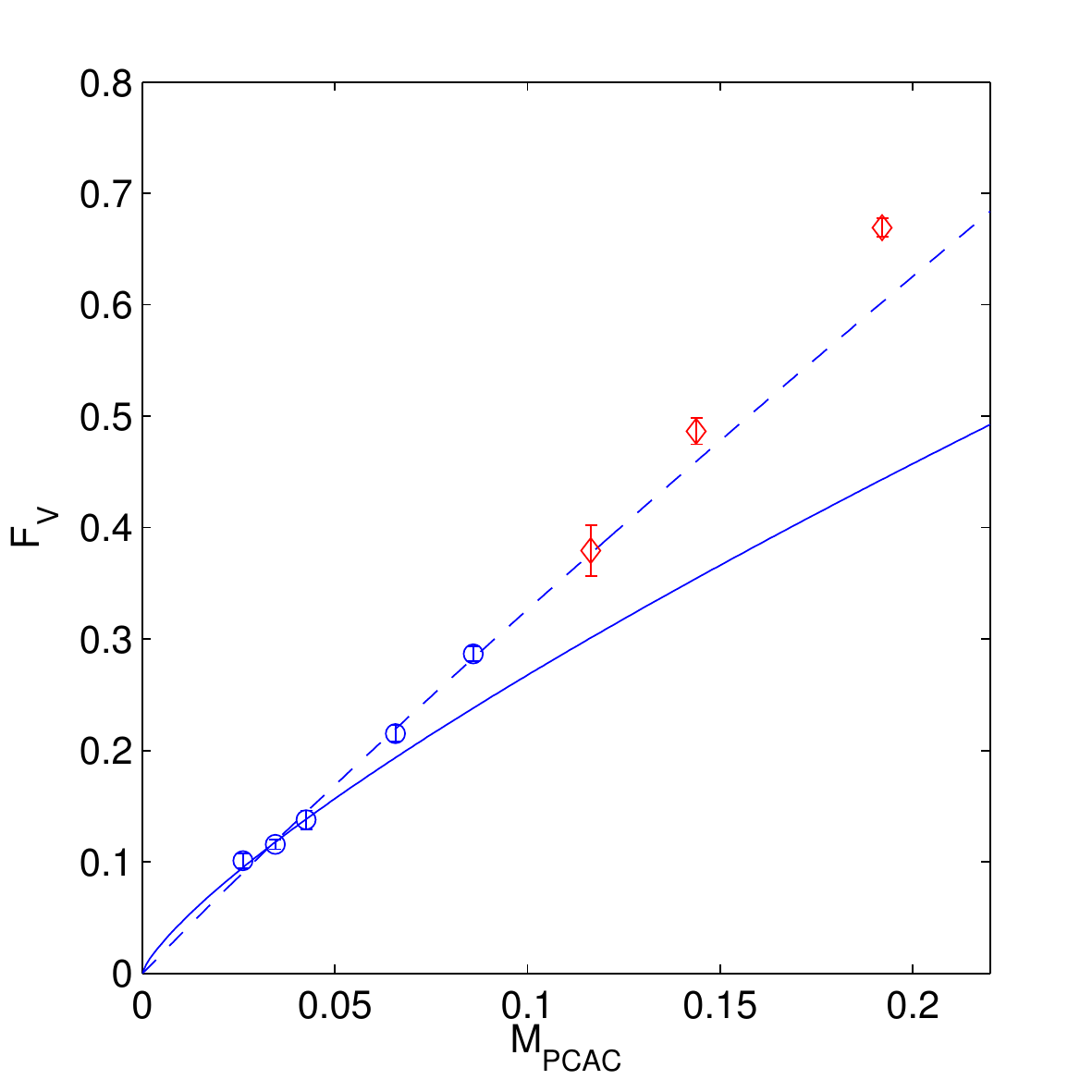} \\
 \vspace{-3mm}
 \includegraphics[width=0.48\textwidth]{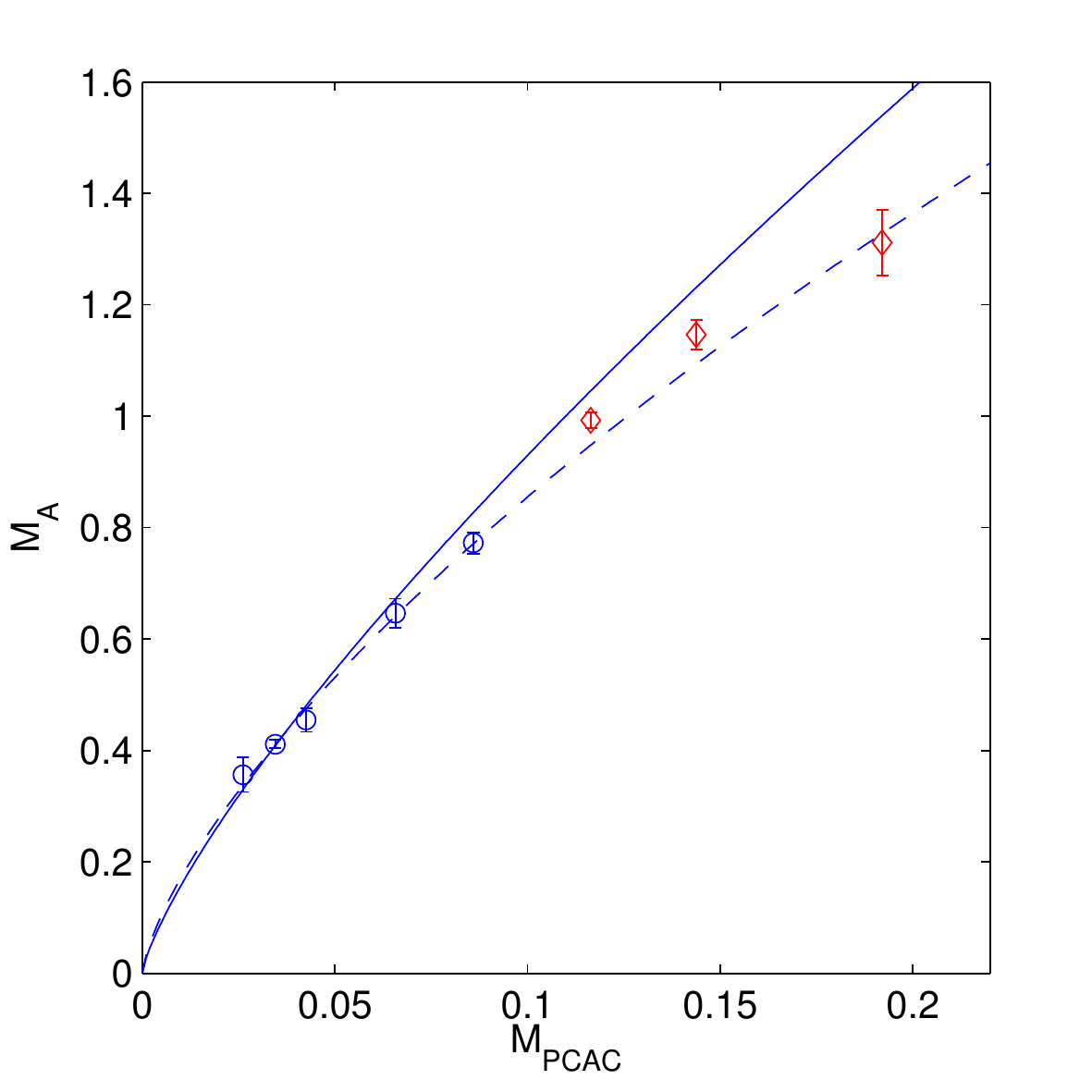}
 \includegraphics[width=0.48\textwidth]{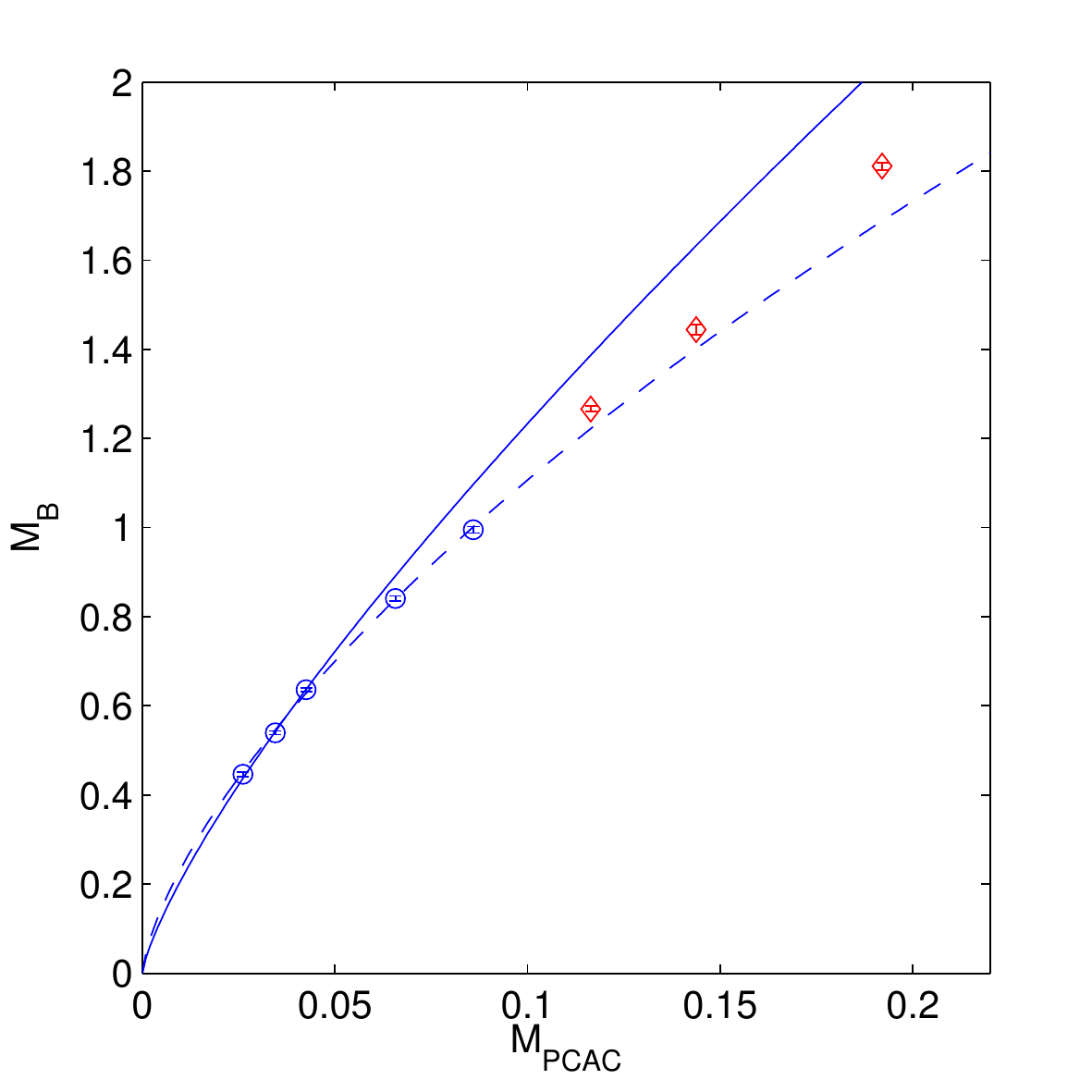} \\
 \vspace{-3mm}
\end{center}
\caption{Combined conformal fits for $\beta=5.4$ with and without subleading corrections. The solid line is the leading order conformal fit to the three lightest points in all channels. The dashed line is the conformal fit, including subleading corrections, to the five lightest points in all channels.}
\label{fig:conformal}
\end{figure}

In a conformal theory all masses $M_x$ and decay constants $F_x$ scale in the same way as a function of the quark mass $m$, and they all vanish in the chiral limit. The scaling behaviour is described by a power law \cite{DelDebbio:2010hu,DelDebbio:2010jy,DelDebbio:2010ze}:
\begin{align}
\begin{split}
 M_x &= A_x m^\frac{1}{1+\gamma}
      + \color{red}\tilde{A}_x m^\frac{1}{1+\alpha_x} \\
 F_x &= B_x m^\frac{1}{1+\gamma}
      + \color{red}\tilde{B}_x m^\frac{1}{1+\beta_x}
\end{split}
\end{align}
Here $x$ is a specific channel and $\gamma$ is the universal scaling exponent (the anomalous dimension of the mass). The second term in each expression is a sub-leading correction, where one allows for a different scaling exponent ($\alpha_x$ and $\beta_x$) in all channels.

In figure \ref{fig:conformal} we show six different channels fitted simultaneously to the conformal scaling behaviour. On the last plot $M_B$ denotes the spin-1/2 baryon. The solid lines are the leading order (LO) conformal fit to the three lightest points in all channels. The dashed lines are the conformal fit, including subleading corrections, to the five lightest points. From the fits we observe that the subleading corrections are more important for the decay constants, compared to the masses.
\begin{table}[h]
\begin{center}
\begin{tabular}{l|cccc}
  & dofs & \# of fitted parameters & $\chi^2$/dof & $\gamma$ \\
  \hline
  LO              & 11 & 7  & 2.9 & 0.29  \\
  LO + subleading & 11 & 19 & 2.3 & 0.23
 \end{tabular}
 \caption{Parameters and results for the conformal fits.}
 \label{tab:conformal}
\end{center}
\vspace{-5mm}
\end{table}
The results of the conformal fits can be found in Table~\ref{tab:conformal}. In both cases $\chi^2$/dof still shows some tension for the fit, similar to the other case considered. We also see that the best fit value for the anomalous dimension $\gamma$ changes a bit when one includes the subleading corrections.

\section{Conclusions}
\vspace{-3mm}
We studied the chiral behaviour of the SU(3) sextet model by comparing its spectrum to the predictions for a chirally broken theory and to those of an IR conformal theory. With our present data both scenarios are still viable since none of the fits we tried are good enough to draw definitive conclusions. We only fitted the spectrum for one of the lattice spacings ($\beta=5.4$) because some systematic uncertainties at $\beta=5.4$ need to be removed before repeating the study on this finer lattice. However, the second lattice spacing already shows the same qualitative behaviour as the one analysed here. To distinguish between the chiral and conformal scenario, consistent results from both lattice spacings are needed. Lighter quark masses are also required. If the theory is chirally broken, at lighter masses we should see an actual splitting between the vector and pseudoscalar mass. On the other hand, if the theory is conformal, with lighter masses we should be able obtain a good fit using the expected LO scaling behaviour. Regardless of the answer, the model is important for our understanding of composite dynamics.

\acknowledgments
\vspace{-2mm}
The CP$^3$-Origins centre is partially funded by the Danish National Research Foundation, grant number DNRF90. This work was supported by a Lundbeck Foundation Fellowship. The computing facilities were provided by the DeIC National HPC Centre.

\end{document}